# Human reliability assessment method applied to investigate human factors in NDT:
## The case of the interpretation of radiograms in the French nuclear sector

Justin LAROUZEE[1], Etienne MARTIN[2], Pierre CALMON[3]

[1] MINES ParisTech, CRC, 06904 Sophia Antipolis France
[2] EDF – DIPNN – Direction Industrielle, 2 rue Ampère 93200 Saint-Denis, France
[3] CEA LIST, Centre de Saclay Bât. 565, F-91191 Gif-sur-Yvette, France

*Abstract*:

This communication reports on a study carried out in the context of the collaborative FOEHN project (Human and Organizational Factors in Non-Destructive Evaluation) supported by the French National Research Agency. The motivation of this project comes from the observation that human and Organizational factors (HOF) are not sufficiently considered by the NDT community. Its goal is to analyse and model the influence of the HOF on selected cases of study in the perspective of a better evaluation of the performance of inspections.

The communication is focused on a radiographic test (RT) case of study in which it appeared that several successive inspections had failed to detect an existing in-service defect. The analysis and modelling of HOF related to interpretation of films has been achieved in the framework of the CREAM (Cognitive and Reliability and Error Analysis Method). A survey has been conducted during the training and the maintaining of the proficiency of NDT (Non Destructive Testing) operators. This was followed by a non-participant observation of operators on site and several individual interviews including a sample of people covering the main organizational and hierarchical roles (eg. project management, management, operations, invigilation).

The exchange with the HOF experts resulted in a hierarchical analysis of "radiogram interpretation" tasks (31 sub-tasks) and a list of contextual and organizational factors that may affect the performance of interpretation of films by the operator. From such a description the CREAM method allows to determine critical tasks and probability of "errors" linked to a limited set of "Common Performance Conditions" (CPC). The first conclusions of this study are that the model CREAM seems well-adapted to the estimation of the impact of HOF on NDT performances. The next phases should be to apply it to other tasks (here only radiograph interpretation) and techniques. The expected benefit of this study is to provide tools for the evaluation and optimisation of NDT implementation.

# 1. Introduction

This communication reports on a study carried out in the context of the collaborative FOEHN project (Human and Organizational Factors in Non-Destructive Evaluation) supported by the French National Agency. The purpose of this project is to assess if and how Human and Organizational Factors (HOF) could be considered in the evaluation of a Probability of Detection (POD), an indicator describing the inspection efficiency. The goal is to analyse and model the influence of the HOF on selected case studies and field observation in the perspective of providing tools for a better evaluation of the performance of inspections.

The communication is focused on Radiographic Testing (RT) in the Nuclear Power Industry. We first present an overview of NDT implementation (section 2). Then, the choice of focusing on RT and especially on the interpretation phase is explained and motivated after a case study in which it appeared that several successive inspections had failed to detect an existing in-service defect (section 3). The analysis and modelling of HOF related to interpretation of radiographs has been achieved using an adapted CREAM (Cognitive and Reliability and Error Analysis Method) framework (section 4). As a result, adapted CREAM methodology allows critical tasks and Human Error Probability (HEP) to be determined (section 5). The first conclusion of this study is that CREAM seems well-adapted to the estimation of the impact of HOF on NDT performances. The next phases should allow it to be applied to other tasks (so far, only radiograph interpretation has been studied), other NDT techniques and ultimately other industrial domains.

# 2. Overview of NDT implementation

NDT plays a vital role in ensuring the safety of Nuclear Power Plants (NPPs) operations. NDT in the context of In Service Inspection (ISI) contributes to the nuclear safety as an essential element of the Defense in Depth concept. NDT is devoted to the research of material degradations and defects such as corrosion or cracking.. The Nuclear industry has devoted considerable attention to optimizing the inspection process for Nuclear Power Plant components. As a result, there has been significant progress in improving NDE reliability over the past two decades through the development of rigorous qualification processes, stand alone and usable procedures, training files and the certification of experienced personnel.

It is well established from cross-industry trials and experience that the reliability of NDT inspection can be significantly affected by human performance issues (e.g. Kettunen & Norros, 1996; Enkvist, *et al.*, 1999; Bertović *et al.*, 2011, 2012; Bertović, 2016; Cumblidge *et al.*, 2017; EPRI, 2017, 2019). Examples of major trials where human factors on inspection have been assessed include the HSE PANI Project, PISC III in the Nuclear Industry, the US Ageing Aircraft Programme. Although performance demonstration helps to ensure that equipment, procedures and personnel are capable of reliably detecting flaws in a formal testing environment, notable failures have occurred during application in field trials. In each case, the equipment and procedures, while not always optimal, were physically capable of obtaining discernible signals from the flaws. Recent events suggest that robust techniques and qualifications are necessary, but may not be sufficient, to accomplish reliable NDE in the field. The effective application of NDE can be dependent on the personnel performing the examination, the design of the task, along with the environmental and organizational conditions within which personnel carry out the task. Furthermore, there is recognition that the performance demonstration environment is very different to working conditions in the field.

Today, NDT operators are subject to extensive preventive measures to improve the safety and reliability of inspections and prevent a catastrophic disaster. How can operators be best prepared for efficient, effective, and satisfying NDT?

The importance of human and organizational factors (HOF) in achieving the safety objectives related to nuclear facilities is unanimously accepted by all stakeholders involved in the design, construction and operation or decommissioning of these facilities. However, beyond this recognition in principle, it is clear that consideration of these factors is still perfectible. They are not systematically integrated into engineering practices. For example, 'human error' is often put forward as a cause in the attempt to explain malfunctions in terms of non-detection of defects. The operator inevitably plays a role in this gap, however involuntary human error (resulting from the psychological and physiological processes involved in the perception, understanding of work situations, decision-making processes, …) and the transgression (also called 'violation') which consists in conscientiously failing to apply a rule or a procedure (Reason, 1990) is often confused.

To ensure an optimal NDT implementation at the first attempt, it's important to know the context and the relationship that exist between the different stakeholders (Regulatory – Utility - Vendor) and the area and responsibility of the Independent Qualification Body (IQB) for the qualification of the NDT System and the RTPO/PCB (Recognized Tierce Party Organisation / Personal Certification Body) for the certification of the personnel. At the beginning, all parties must know the strengths and the difficulties to inspect with a qualified NDT System, even if the upstream actors believe that the last but not least barrier (NDT) of the structural integrity can address all situations. To implement an NDT system in an optimal situation, it's important to put the emphasis on:
- Qualification of the procedure of NDT System by an IQB,
- Certification of the personnel by an RTPO / PCB,
- Responsibilities of the vendor to maintain the proficiency of the personnel.

After that, many factors can influence the quality and successful completion of examination. Operational Experience (OE) from in-service inspections of nuclear power plants (NPP) includes cases where oversights or poor planning practices resulted in problems with the application of qualified NDE procedures. These problems result in degraded examination performance such as incomplete examination coverage, mistaken identification of service-induced defects where none exist, or failure to identify existing service-induced defects.

A Utility should ensure that the vendor personnel has received procedure training which addresses the examination scope, the specific NDT procedures, equipment, and software that are to be used:
- During the pre-job briefing, the utility should discuss the verification of surface condition and encourage prompt reporting of any examination limitations or concerns held by the examiner
- Vendor should perform a detailed readiness review in advance of the outage. Utility involvement in this review is highly recommended.

## 3. Situation: Case Analysis of RT implementation

Despite these recommendations, at the end it is possible to encounter a problem. One specific case has been used to first analyse where things can go wrong and what kind of HOF method would be needed to improve the understanding and prevention of such events. In this section, we first present the description of the non-detection of an in-service defect (3.1). Then we present our analysis of the different failures during the inspections (3.2). Finally, we propose an approach to determine the reasons of these failures taking into account the HOF (3.3).

### 3.1 Case description

A bimetallic butt-weld end connection of valve (ø 3 "x 2.77 mm), replaced in 2011, is the subject of an in-service radiographic follow-up examination (48-month periodicity). The sought degradation is mechanical vibratory fatigue cracking. In 2016, 13 months after the last gamma ray inspection, a leak under insulation led the NPP to detect a jet of steam at the bimetallic weld.

The weld was gamma-rayed 3 times, interpreted 6 times by 5 different SQEP (Suitably Qualified Experimented Personnel) but no linear image has ever been recorded. The rereading of all the radiographs made on this weld (2012, 2014 and 2015) revealed a 70 mm image at the edge of a root, stainless steel side, with circumferential orientation.

### 3.2 Problem

A first analysis showed different failures to respect the Inspection Procedure. First, as weakness of the *vendor*; according to the procedure, the interpretation and analysis of the manufacturing radiographs should have been able to guide the analysis towards a degradation that appeared in service, of the crack type , but they were not requested from the NPP by the different NDE providers. Second, as weakness of the *invigilators*; with the feedback of the qualification, the utility can put special emphasis on some points of invigilation. In readiness of an invigilation of the provider, the utility personnel should have a thorough understanding of the examination procedure, both for preparation and for oversight during the examination process in advance of the outage. Likewise, utility personnel should be aware of the procedure's limitations and its impact on flaw assessment and code acceptability of an examination.

But even more than identifying the specific causes of the case, the event appeared to us as an opportunity to undertake an in-depth study of the human contribution to the reliability of NDT. The objective was to try to understand how, in spite of the proficiency of the operators, the organization could fail so that 3 inspectors during 3 outages had not seen the image of the defect on the radiographs. We encountered a refusal from the operators involved to respond to an interview (organised by an external auditor). Some even argued that they were not involved in this NDT case. This reluctance clearly indicated the importance but also the passionate nature of the subject within the organizations concerned. That's why it was decided to address the HOF subject on the basis of observation of nominal work situations and as part of a multi-partner project.

### 3.3 Resolution

When addressing the uncertainties (operator errors, misunderstandings, oversight mistakes) related to humans, one can distinguish two approaches: the standard mathematical approaches (probability theory, fuzzy logic or data-driven models), or human reliability and performance and its factors of influence. The latter is also called Human Reliability Analysis (HRA) methods, most of them are domain-specific but few can be adapted (Bell and Holroyd, 2009).

In order to account for the influencing factors of the activity and to develop a tool capable of estimating a (qualitative) probability of human error, we decided to adapt a proven HRA method to the NDT context. To avoid the shortcomings of the so-called "1st generation" HRA-methods[1] we decided to use a second generation HRA-method: Cognitive Reliability and Error Analysis Method (CREAM)[2]. This method was chosen for adaptation and utilization because it considers the influence of working environment on human's performance, it can be applied by a person having a good knowledge of the process and provides qualitative results, expressed in an easy to interpret and use.

---

[1] Their objective is to find the *Human error probability* (HEP) based on the skill- and rule-based level of human action which neglects the context, errors of commission, etc. For a list of 1st generation shortcomings, see Hollnagel, 1998.

## 4. Applying CREAM to the radiograph interpretation

CREAM has been used within the safety evaluation process of several critical systems (e.g. trains, NPPs, space-shuttles, see Gertman & Blackman, 2001), it was first presented by Erik Hollnagel in 1998, and its core idea is that human error is shaped by both the operational context and human cognitive and biological nature. In this section, we first present the fundamentals and philosophy of CREAM (4.1). Then we present the adaptations that have been made to CREAM within the FOEHN project (4.2). Finally, we propose a (simplified) example of implementation (4.3).

### 4.1 Presentation of the method

The strength of CREAM lies in its underlying model rooted in cognitive and behavior sciences, which is one of the major characteristics of an advanced HRA method as listed by Mosleh (2001). CREAM can be used in both *retrospective* and *prospective* ways. As our objective was to estimate human related risks of failure in the NDT process, we adapted the *prospective* mode. Each mode exists in a *basic* and an *extended* version. The core of CREAM is that human error is shaped by the context of the task and thus, not stochastic. The method identifies nine common performance conditions (CPC) that together provides a structured and comprehensive way of characterizing the conditions under which activity is expected to take place. The *basic* version consists in the study of each CPC for a given task that allows determining a *control mode* reflecting the characteristics of the different conditions. Based on the context description the CPCs provide, the control mode provides an indication of the degree of control that an operator or one team has over the situation. There are four control modes defined by Hollnagel: scrambled, opportunistic, tactical and strategic (the error probability being gradually reduced).

The *control mode* depends on the combined characteristic of all CPCs, expressed as their combined score, derived by counting the occurrences where a CPC is expected to reduce, have no significant effect on, or to improve performance reliability. Hollnagel thus ends with a matrix he calls COCOM (for Cognitive Control Model, see Figure 1).

---

[2] CREAM is one of the most used HRA methods, other well none methods include Technique for Human Error Rate Prediction (THERP) or Human Error Assessment and Reduction Technique (HEART).

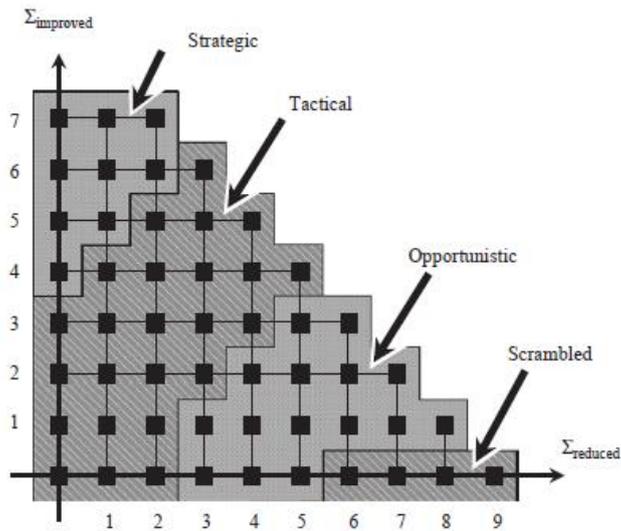

*Figure 1: COCOM links the CPCs states and cognitive control modes (after Hollnalgel, 1998)*

Ultimately, each control mode is linked to a range of probabilities of incorrectly performing the task (see Table 1). It is interesting to note that COCOM has been empirically validated by independent studies (see for example Stanton *et al.,* 2001), which gives it great value.

*Table 1 : Control modes and error probability intervals (after Hollnagel, 1998)*

Control modes and probability interval

| Control mode | Probability interval |
|---|---|
| Strategic | $0.00005 < P < 0.01$ |
| Tactic | $0.001 < P < 0.1$ |
| Opportunistic | $0.01 < P < 0.5$ |
| Scrambled | $0.1 < P < 1.0$ |

In the basic mode, the analysis is not focused on the cognitive human activities involved and their links to human-errors but on the working environment (physical and organizational). Thus, it can be considered as an initial screening of the operator's actions and a way to produce an overall assessment of the performance reliability. The *extended* version further develops the results of the basic one and distinguishes the most probable failures mode and probability for each segment of an analyzed task. It consists of the following steps: (1) identify the cognitive activities of the tasks to build a cognitive profile, (2) identify a most likely cognitive function failure for each identified cognitive activity, and (3) determine the probability for each identified cognitive function failure.

There are 13 Generic Failure Types (GFTs) in observation, interpretation, planning, and execution, and each has a nominal failure probability (see Table 2).

*Table 2 : Nominal values for generic cognitive failure types* (after Hollnagel, 1998)

| Cognitive function | Generic failure type | Basic value |
|---|---|---|
| Observation | O1. Wrong object observed | 0.001 |
| | O2. Wrong identification | 0.007 |
| | O3. Observation not made | 0.007 |
| Interpretation | I1. Faulty diagnosis | 0.02 |
| | I2. Decision error | 0.01 |
| | I3. Delayed interpretation | 0.01 |
| Planning | P1. Priority error | 0.01 |
| | P2. Inadequate plan | 0.01 |
| Execution | E1. Action of wrong type | 0.003 |
| | E2. Action at wrong time | 0.003 |
| | E3. Action on wrong object | 0.0005 |
| | E4. Action out of sequence | 0.003 |
| | E5. Missed action | 0.003 |

The causes of these failures can fall into three categories: those related to the individual, the technology and/or the organization. The category of causes associated with the individual is composed of three groups of antecedents. The first concerns the specific cognitive functions of the individual and the other two concern more general functions: temporary and permanent functions. The specific cognitive functions make it possible to explain the observed action. To categorize the functions that are the basis of thinking and decision making, Hollnagel differentiates analysis from synthesis.

There are thus four subgroups, two for the analysis part: the observation subgroup (missed observations, false observations and identification errors) and the interpretation subgroup (errors of diagnosis, reasoning, decision, late interpretation, incorrect prediction); two for the synthesis part: the planning subgroup (inadequate plans and priority errors) and the execution subgroup (already represented by the category phenotypes). Temporary functions associated with the person describe the psychological, physical and emotional states of the person that may be of limited duration (memory problems, fear, distraction, fatigue, inattention, stress and performance variability). Finally, the permanent functions associated with the person describe the constant characteristics of the individual (disability, style and cognitive bias). The second category, technology-related causes, covers technical malfunctions and defects, procedural inadequacies, and interface problems. This category consists of four groups of history: equipment (hardware and software failure); procedures (inadequate procedures); temporary interface problems (access limitation, ambiguous and/or incomplete information); and permanent interface problems (access problems, labeling/naming problems). The third and last category, that of causes associated with the organization (including the work environment), is composed of five groups of antecedents: communication (communication problem, lack of information); organization (maintenance problem, quality control, management, social pressure, etc.); training (insufficient skills and/or knowledge); environmental conditions (temperature, noise, etc.); and working conditions (excessive demand, irregular hours, etc.).

On the basis of these elements and as mentioned above, CREAM can be applied to different areas and applications. This is mainly due to a certain "genericity" of its classification scheme. Nevertheless, Hollnagel points out that this scheme can be modified and adapted to a certain application or field by changing certain categories. In the following, we summarize the main modifications and adaptations we have made.

## 4.2 Adaptation of the method

The objective of our work was to obtain a quantified probability of human error; thus, the basic version of CREAM wasn't relevant for us and we naturally oriented our choice on the extended version. In order to use the extended version, we had to address two main tasks: first we had to construct an event sequence for the task chosen for analysis (i.e. interpretation of radiographic films); second, we had to adapt the CPCs to the specific context of NDT.

Through the project, two weeks of non-participative observations in two different NPPs were conducted. The field observations were completed by twenty interviews conducted with RT technicians, engineers and management. Also, five focus-groups were organized and held with RT experts. The raw data collected was used to build a hierarchic task analysis (4.2.1) and to adapt the CPCs (4.2.2).

### *4.2.1 Construction of an event sequence*

The aim of this first step is to create an event sequence providing a detailed analysis of the studied activity, which will be a basis for the subsequent steps. We use the Hierarchical Task Analysis (HTA) method to do so (Annett, 2003). Based on the non-participant observations conducted on site, we built a list of main tasks that constitute the activity and decomposed them until the sub-steps represented elementary actions.

At a high level, the activity of interpreting a radiograph is divided into four macro sub-activities:

1. **Preparing the workspace** (11 sub-tasks): after the operator has entered the interpreting room, this consists of switching on the devices, checking their conformity, making the first adjustments and/or calibrations.
2. **Preparing the interpretation** (9 sub-tasks): once installed in a prepared space, the operator must check that he or she has the necessary small equipment (ruler, blank examination report, white gloves, etc.) and, above all, that the procedure(s) he or she will have to apply is (are) available and within reach.
3. **Verifying the quality/conformity of the radiographic film** (18 sub-tasks): at this stage the actual handling of the radiographic films begins. A series of tasks must be carried out to ensure that the films meet the quality criteria and comply with the procedure to be applied.
4. **Interpreting the radiographic film** (22 sub-tasks): this last step represents the controller's core business. It consists in methodologically identifying and characterizing (size, location) any significant or suspicious contrast difference the radiographs.

As it would be too long to detail in this article the 60 sub-tasks of interpretation formalized in the HTA, in the rest of this article, we will only detail the step 3 (verifying the quality/conformity of the radiograph) as a support to the demonstration.

### *4.2.2 Selection and adaptation for relevant common performance conditions*

Observation of professionals and interviews with experts in the field was also used to adapt the CPCs of the original CREAM. This adaptation is necessary to make the estimation of the probability of error as accurate as possible, but also to make the adapted method self-supporting. Thus, the vocabulary used to name and define the CPCs must be as explicit and precise as possible.

Table 4 below presents a summary of the adapted CPCs. These CPCs, which are intended to be limited in number, are capable of influencing the performance of the interpreter. The ambition is to be able to describe, on the basis of a unique and defined framework, any working situation. The general principles described earlier (4.1) remain valid: each CPC can thus have a positive or negative influence on the human performance depending on their state.

*Table 3 : **Adapted CPCs for the context of RT-NDE.** Both their name, description and order have been changed from the original CREAM. Eligible states are the states one can set any given CPC and the correlated influence on the operator's cognitive control thanks to the COCOM (qualitatively "+" means improved cognitive control, "0" means no significant effect, "-" means decreased cognitive control).*

| # | CPC | DESCRIPTION / ANALYSIS ELEMENTS |
|---|---|---|
| 1 | PROCEDURES & TECHNICAL DOCUMENTATION *, History of previous exams, examination protocols, charts, abacus)* | Available, displayed, up to date (indexes, revisions), manufacturing films (archiving), end of manufacturing report, design plan, welding records, etc. **Eligible states:** Appropriate (+), Acceptable (0), Inappropriate (-) |
| 2 | NUMBER OF SIMULTANEOUS OBJECTIVES | Number of welds to be interpreted (possibly number of radiographic films per weld). Type of problem (need for additional exposures, meeting). Acceptance or not of radiographics films (films conformity). Type of weld (attention when in need to interpret different welds), the diversity of the controlled areas, their geometry. This is related to the organization of work since it is tried to distribute the workload on available interpreters so that they neither use a single procedure overnight nor have change too frequently. **Eligible states:** Less than capacity (0), At capacity (0), More than capacity (-) |
| 3 | LOCAL CONDITIONS OF INTERPRETATION | Interpretation room condition (dark, quiet, wall color, 0-10 LUX, not used as a checkroom, twinned with the lab, multiple interpreters in parallel). Sufficient space to deposit films, to fill out examination reports. Adequate temperature of the room (the use of several illuminators in a small room requires the use of air conditioning to lower the temperature, otherwise the illuminators might break down. Meanwhile several illuminators generate a lot of noise from their fans, which is tiring in the long run). **Eligible states:** Advantageous (+), Compatible (0), Incompatible (-) |
| 4 | AVAILABLE TIME | Number of radiographs to be interpreted, phase of the shutdown. Need to catch up with delays (end of intervention on critical path). Possible pressure to communicate the results quickly and to be able to release the material (sequence of phases). Fatigue. **Eligible states:** Adequate (+), Temporarily inadequate (0), Continually inadequate (-) |
| 5 | QUALITY OF THE HARDWARE | Collective equipment: illuminators (with foot pedal to avoid flashes) and densitometers, type of illuminator (with adjustable iris, rectangular, LED, etc.). Individual equipment: gloves, ruler, pencil, reading table... **Eligible states:** Adequate, verified (0), Satisfactory (0), Inadequate (-) |
| 6 | TRAINING AND EXPERIENCE | Level of education (level of knowledge in the broadest sense; it is not the level of education that makes the interpreter), years of experience, arrangements, and frequency of skill maintenance. Duration of the companionship **Eligible states:** Adequate training, experienced (+), Adequate training, little experience (0), Inadequate (-) |
| 7 | EFFECTIVENESS OF COLLABORATION / COMMUNICATION | Relations between NPP, internal engineering, technical assistants, and the provider. Negotiations on position, exposure. Collaboration or competition between teams (preparation, exposure, lab, interpreter). Loss of concentration due to the presence of clients in the interpretation room (right behind the interpreter implying a form of pressure). Exercise of free will (avoid the phenomenon of copying/adopting the opinion of another person). **Eligible states:** Very efficient (+), Efficient (0), Inefficient (-), Undesirable (-) |

| 8 | TIME OF DAY<br>*(day = 8am to 8pm)*<br>PERIOD OF THE WEEK<br>*(middle = Tuesday, Wednesday, Thursday)* | Favorable end of night (calm, less co-activity), but also more fatigue (monitor the accumulation of weekends worked and the weekly workload). Beginning and end of weeks at risk: long distance driving (from home to work) and/or workload accumulation.<br><br>**Eligible states:** Day, mid-week (0), Day, early or late week (0), Night, mid-week (+), Night, beginning or end of week (-) |
|---|---|---|

The adaptation of CREAM to make it understandable, useful and applicable by NDT professionals regardless of their level of knowledge in the field of human factors became one of the main focus of the research project. Thus, the result section (5.) presents the output of the simplified CREAM methodology and its interest for the NDT community, regardless of other developments that have been made in the extended CREAM methodology.

# 5. Simplified results and discussion

In this section, we present some macroscopic results of the use of the CREAM method adapted to RT tests (5.1). It would take too long to detail the fine steps that lead to error probabilities. This aspect will be treated in the discussion (5.2). However, it is very interesting to note that without going into too much detail, the exercise of adapting and applying the CREAM method provides lines of thought that are probably new for the profession.

## 5.1 Cognitive demand profile

In order to understand the cognitive activities mobilized in the interpretation activity and consequently which types of failures (errors) are most likely to occur, the CREAM method proposes to establish a cognitive load profile. Thus, 15 elementary mental (cognitive) activities are considered: coordinate, communicate, compare, diagnose, evaluate, execute, identify, maintain, monitor, observe, plan, record, regulate, scan, and verify. The CREAM method proposes to group them together to distinguish four main cognitive functions: observe, interpret, plan, and execute.

To construct a cognitive demand profile, each item in the task analysis must be assigned to one of the four main cognitive functions. In our case, this assignment was performed by several experts in the field supervised by the author in specific group sessions. A sample result of this step for the step #3 "check the quality of the film" is shown in the table 4 bellow:

*Table 4 : extracted from the hierarchical task analysis. The activity is finely detailed and decomposed so that it can be the subject of cognitive analyses to determine the probability of human error. With each sub-task is associated a cognitive function (CF) and the most probable cognitive function failure (CFF) is estimated and associated with a cognitive failure probability (CFP).*

| #3 : Check the quality of the film | | | | |
|---|---|---|---|---|
| # | Sub-task | CF | CFF | CFP |
| 3.1 | Grabbing a radiograph | | | |
| 3.1.1 | Open paper folder | Planning | P1 | 1,00E-02 |
| 3.1.2 | Remove paper spacer | Planning | P1 | 1,00E-02 |
| 3.2 | Check the adequacy of the luminous marking, the lead marker with the operating conditions of exposure | Observation | O2 | 7,00E-02 |
| | | Planning | P1 | 1,00E-02 |
| 3.3 | Checking the quality of radiographs | | | |
| 3.3.1 | Checking the general condition of the radiograph | Observation | O2 | 7,00E-02 |
| | | Execution | E5 | 3,00E-02 |
| 3.3.2 | Determining the area to be interpreted | Planning | P2 | 1,00E-02 |
| | | Execution | E3 | 5,00E-04 |
| 3.3.3 | Check the completeness of the area to be interpreted | Observation | O2 | 7,00E-02 |
| | | Planning | P1 | 1,00E-02 |
| | | Execution | E3 | 5,00E-04 |
| 3.3.4 | Check radiograph density | | | |
| 3.3.4.1 | Use densitometer | Planning | P2 | 1,00E-02 |
| 3.3.4.2 | Survey the density in accordance with the procedure | Planning | P2 | 1,00E-02 |
| | | Execution | E1 | 3,00E-03 |
| 3.3.4.3 | Check that the density meets the requirements of the procedure | Observation | O2 | 7,00E-02 |
| | | Planning | P2 | 1,00E-02 |
| | | Execution | E1 | 3,00E-03 |
| 3.3.4.4 | Record the density values of the radiograph in the examination report. | Execution | E4 | 3,00E-03 |
| 3.3.4.5 | Check the difference in density at the same point of the two films in the radiograph | Observation | O1 | 1,00E-03 |
| | | Planning | P2 | 1,00E-02 |
| | | Execution | E5 | 3,00E-02 |
| 3.3.5 | Checking the conformity of the radiograph image quality indicator | Observer | O3 | 7,00E-02 |
| | | Planning | P1 | 1,00E-02 |
| | | Execution | E4 | 3,00E-03 |
| 3.3.5.1 | Record the Image Quality Indicator (IQI) in the examination report | Execution | E5 | 3,00E-02 |
| 3.3.6 | Check the conformity of the numbered strip (Presence of indicator, form step (distance in marks and between indicators), position, and parallelism) | Observation | O3 | 7,00E-02 |
| | | Planning | P2 | 1,00E-02 |
| | | Execution | E5 | 3,00E-02 |
| 3.3.7 | Check that the entire weld has been radiographed (overlap) | Observation | O1 | 1,00E-03 |
| | | Planning | P2 | 1,00E-02 |
| | | Execution | E5 | 3,00E-02 |

From equivalent analysis and characterization on the four steps defined earlier (4.2.1), an overall cognitive demand profile can be built and presented in the form of histograms (Figure 3) :

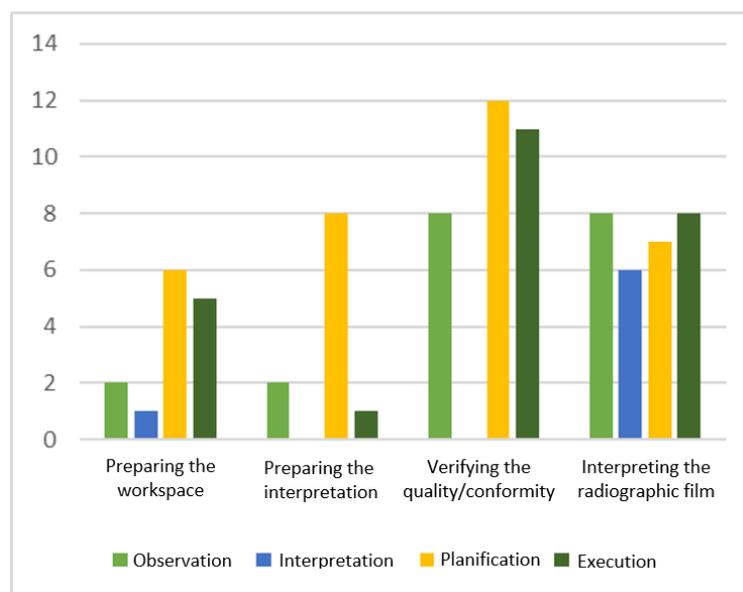

*Figure 2 : **Cognitive demand profile** for each of the four steps composing the film interpretation*

Such a cognitive demand profile shows that the activity of radiograph interpretation only marginally mobilizes the cognitive function of interpretation (see Figure 4). If such a result remains too generic to initiate targeted and pragmatic error reduction measures, its degree of genericity and its counter-intuitive character makes it useful for communication, training or simply to discuss of the human and organizational factors in NDT.

Fortunately, if it is and will probably ever remain uncertain how to objectively secure a purely mental operation such as interpretation, proven countermeasures do exist for planning, execution or observation operations.

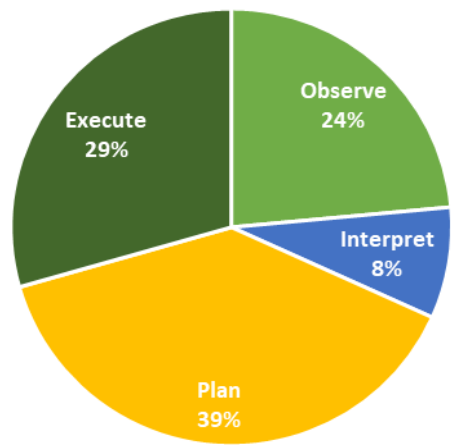

*Figure 3 : Cognitive demand profile related to film interpretation* (for all the 60 sub-tasks)

## 5.2 Discussion

Despite the real value of the studied case and the field data collected during this research project, we must agree with Librizzi & Zio (2006) when they state that a critical issue in the assessment of human performance (or error) is the quantification of the expert judgments with respect to the context in which the human action takes place. Due to the intrinsic difficulty of this evaluation and to the lack historical and structured data (in other words a strong return on experience), the choice of a CPC state remains subjective and affected by uncertainty.

There are some attempts in the literature to improve the human error quantification through the combination of CREAM with fuzzy logic (Konstandinidou *et al*., 2006) or Bayesian networks (Kim *et al*., 2006). But as far as we are concerned, the main challenge for the NDT community may not be of such an advanced level. Indeed, the campaign of field surveys and the numerous individual and collective semi-directive interviews with a sample of people covering the main organizational and hierarchical roles (e.g. project management, management, operations, supervision) has shown that the concepts and issues related to human and organizational factors in the profession still lacks appropriation. In particular, we noted that the distinction between human and organizational factors is not always well known, which contributes to neglecting the impact of certain contextual aspects on performance whatever the methodological approach may be. Moreover, we noted that the human contribution is generally perceived negatively (e.g. from the perspective of violation). Thus, instead of investing on the quantification of human error probabilities, the partners of FOEHN project are willing to stimulate reflections and debates in the field with the initiation of a national scale taskforce under the aegis of the French Personal Certification Body: COFREND (Confédération Française des Essais Non Destructifs).

# 6. Conclusion

Nondestructive evaluation (NDE) in the Nuclear Power Industry represents one of the most extreme and demanding work roles and environments for skilled personnel. Examiners may perform their jobs under conditions of high physical and psychological stress from many sources (including heat, radiation exposure, time pressure, noise, etc.). It is important that the Plant Owner works closely with the examination vendor and personnel to ensure reliable examinations. Good communication, collaboration, and leadership between the Plant Owner and the vendor are key elements in the quest for the most reliable examinations. Examination personnel need to comprehend the entire examination activity and realize that it must be carried out with sufficient time and rigor to thoroughly assess component condition and identify defects. Thus, the so called "safety culture" amongst NDE practitioners should be monitored and periodically reinforced.

Based on the information collected during the interviews, the CREAM method was adapted to the NDE context and deployed to analyze the film interpretation phase (critical component of the NDE process). The method was adapted in terms of indicators labels and definitions so it can be used as a stand-alone analysis method for retrospective or prospective analysis. In fact, to help NDE professional that might not be aware of the concepts and methods of the HOF field we provide:

- A very detailed hierarchical analysis of the tasks dissecting the interpreting activity into 60 subtasks to which were associated a cognitive function (observe, interpret, plan or execute) as well as, by concatenation of expert opinions, the most likely failure of this function.
- An adapted set of CPCs to the specific context of an interpretation room (in terms of definitions and admissible states).

Thus, the adapted CREAM was effectively used by the Licensees or Plant Nuclear Owners to facilitate the preparation of work sites and/or to help the invigilators focus their controls on certain specific points related to organization (to date, within EDF, there have been two internal working groups using the CREAM method for the guides of survey and there is a COFREND TV working group). We believe that one of the significant contributions of this transposition of CREAM is to clarify and specify what can be understood as an "organizational factor", as this organizational dimension has remained largely unclear, compared with the human dimension of "human and organizational factors" of safety.

Moreover, the method allows to produce an estimate of the probability of error, depending on the context in which the activity is performed, on the different sequences that compose an interpretation. For this purpose, the cognitive control model proposed by Hollnagel (COCOM) must be used to account for the modalities of task realization and to estimate error probabilities. This represents an interesting path to integrate HOF data in the POD calculation. If the studies carried out during the project do not question the relevance of probabilistic approaches, they lead to the dismissal of any attempt to translate the influence of human performance by the metric currently used to quantify the technical performance of a process (i.e. the size of the largest defect possibly missed). Far from claiming to objectify the human factor on the basis of deterministic criteria, the CREAM methodology should demonstrate that human performance is fundamentally systemic and multifactorial. Further work is necessary to provide a realistic simulation of the work-as-done allowing robust quantitative characterization (Rodat *et al.*, 2017).

We will continue to validate the CREAM method and further expand this model through review of operating experience, observations of NDE in the field and at the different step of development and qualification of an NDT system for the others methods / techniques, and through input from SMEs (Subject Matter Experts ) (e.g. interviews, focus groups, discussions). Once the model is completed, we will seek expert input to prioritize the human factors issues identified. The information gathered in this project will serve as input to the development of a long-term strategy and plan for addressing human performance in NDE.

## 6.1 Références

Annett, J. (2003). Hierarchical task analysis. In E. Hollnagel (Ed.), Handbook of

Bell, J., & Holroyd, J. (2009). Review of human reliability assessment methods. Health & Safety Laboratory, 78.

Bertović, M. (2016). Human factors in non-destructive testing (NDT): risks and challenges of mechanised NDT.

Bertović, M., Fahlbruch, B., Müller, C., Pitkänen, J., Ronneteg, U., Gaal, M., & Schombach, D. (2012, April). Human factors approach to the acquisition and evaluation of NDT data. In 18th World Conference on Nondestructive Testing.

Bertović, M., Gaal, M., Müller, C., & Fahlbruch, B. (2011). Investigating human factors in manual ultrasonic testing: testing the human factors model. Insight-Non-Destructive Testing and Condition Monitoring, 53(12), 673-676.

Cumblidge, S., D'Agostino, A., Morrow, S., Franklin, C., & Hughes, N. (2017). Review of Human Factors Research in Nondestructive Examination. US Nuclear Regulatory Commission - Pacific Northwest National Laboratory) 7th European-American Workshop on Reliability of NDE 2017.

Enkvist, J., Edland, A., & Svenson, O. (1999). Human factors aspects of non-destructive testing in the nuclear power context. A review of research in the field (No. SKI-R-99-8). Swedish Nuclear Power Inspectorate.

EPRI Report: "Human Factors in Nondestructive Evaluation (NDE) A Literature Review and Field Observations" - Ref 3002010462 - Technical Update, November 2017.

EPRI Report: "Human Factors in Nondestructive Evaluation: Manual Ultrasonic Testing" - Ref 3002015981 - Final Report, December 2019.

Gertman, D., Blackman, H. (2001). *Sensitivity analysis case study: incorporating organisational factors in HRA*. Proceeding of the international workshop building the new HRA: errors of commission from research to application, 2001.

Hollnagel, E. (1998). *Cognitive reliability and error analysis method* (CREAM). Elsevier.

Kettunen, J., & Norros, L. (1996). Human and organisational factors influencing the reliability of non-destructive testing. An international literary survey (No. STUK-YTO-TR--103). Finnish Centre for Radiation and Nuclear Safety (STUK).

Konstandinidou, M., Nivolianitou, Z., Kiranoudis, C., & Markatos, N. (2006). A fuzzy modeling application of CREAM methodology for human reliability analysis. *Reliability Engineering & System Safety*, *91*(6), 706-716.

Librizzi, M., & Zio, E. (2006). Sensitivity Analysis of the Weighted Fuzzy CREAM for the Assessment of Human Performance. In ESREL 2006, European Safety and Reliability Conference (pp. 457-464). Estoril, Portugal.

Mosleh A. (2001). On characteristics of advanced HRA models. OECD workshop on errors of commission, Rockville, MD, May 7–9, 2001.

Reason, J. (1990). Human error. Cambridge university press.

Rodat, D., Guibert, F., Dominguez, N., & Calmon, P. (2017, February). Operational NDT simulator, towards human factors integration in simulated probability of detection. In AIP Conference Proceedings (Vol. 1806, No. 1, p. 140004). AIP Publishing LLC.

Stanton, N. A., Ashleigh, M. J., Roberts, A. D., & Xu, F. (2001). Testing Hollnagel's contextual control model: Assessing team behaviour in a human supervisory control task. Journal of Cognitive Ergonomics, 5(1), 21-33.